# Six Fallacies in Substituting Large Language Models for Human Participants

Zhicheng Lin
Department of Psychology, Yonsei University
Department of Psychology, University of Science and Technology of China

**Correspondence**
Zhicheng Lin, Department of Psychology, Yonsei University, Seoul, 03722, Republic of Korea (zhichenglin@gmail.com; X/Twitter: @ZLinPsy)
**Acknowledgments**
I thank Gati Aher, Michael Bernstein, Danica Dillion, Nancy Fulda, Nicholas Laskowski, Paweł Niszczota, Philipp Schoenegger, Lindia Tjuatja, Lukasz Walasek, and David Wingate for comments on early drafts. The writing was supported by the National Key R&D Program of China STI2030 Major Projects (2021ZD0204200), the National Natural Science Foundation of China (32071045), and the Shenzhen Fundamental Research Program (JCYJ20210324134603010). The funders had no role in the decision to publish or in the preparation of the manuscript. I used Claude Sonnet/Opus 4, GPT-4o, and Gemini 2.5 Pro for proofreading the manuscript, following the prompts described at https://www.nature.com/articles/s41551-024-01185-8.

## Abstract

Can AI systems like large language models (LLMs) replace human participants in behavioral and psychological research? Here I critically evaluate the "replacement" perspective and identify six interpretive fallacies that undermine its validity. These fallacies are: (1) equating token prediction with human intelligence, (2) treating LLMs as the average human, (3) interpreting alignment as explanation, (4) anthropomorphizing AI systems, (5) essentializing identities, and (6) substituting model data for human evidence. Each fallacy represents a potential misunderstanding about what LLMs are and what they can tell us about human cognition. The analysis distinguishes levels of similarity between LLMs and humans, particularly functional equivalence (outputs) versus mechanistic equivalence (processes), while highlighting both technical limitations (addressable through engineering) and conceptual limitations (arising from fundamental differences between statistical and biological intelligence). For each fallacy, specific safeguards are provided to guide responsible research practices. Ultimately, the analysis supports conceptualizing LLMs as pragmatic simulation tools—useful for role-play, rapid hypothesis testing, and computational modeling (provided their outputs are validated against human data)—rather than as replacements for human participants. This framework enables researchers to leverage language models productively while respecting the fundamental differences between machine intelligence and human thought.

*Keywords:* generative AI (GenAI), large language models (LLMs), AI participants (AI subjects), silicon sampling, simulation or modeling (modelling), research validity

**Six Fallacies in Substituting Large Language Models for Human Participants**

Collecting data from human participants is resource-intensive and time-consuming. The rise of large language models (LLMs) capable of generating human-like text has sparked substantial interest among psychologists and social scientists (Abdurahman et al., 2025; Demszky et al., 2023; Lin, 2023). In tasks ranging from perception and cognition to language and moral reasoning, LLMs sometimes produce responses that mirror those of average human participants (Aher et al., 2023; Binz & Schulz, 2023a; Dasgupta et al., 2022; Dillion et al., 2023; Hu et al., 2024; Marjieh et al., 2024). As these models are increasingly incorporated into experimental designs and research methodologies, promising instant and tireless responses, a question arises: Could they *ever* stand in for humans in behavioral and psychological research?

Across various fields, proposals have emerged suggesting that LLMs could, to varying degrees, "substitute human participants" in empirical research as "silicon samples" (Sarstedt et al., 2024); "supplant human participants for data collection" in social science as "simulated participants" (Grossmann et al., 2023); serve as "a single participant" and "a proxy for human participants in a certain set of circumstances" in psychological science (Dillion et al., 2023); and "replace humans" in human-centered design to provide "simulated user responses" (Schmidt et al., 2024). Echoing this perspective, commercial entities now peddle synthetic, human-like AI subjects for user and market research—advertising user research "without the users."

While replacement of human participants is not yet mainstream practice, these emerging viewpoints reflect a broader narrative in AI development—one that envisions systems that could "outperform humans at most economically valuable work" (OpenAI, 2024) or "on a vast array of tasks" (Bengio, 2024). While AI researchers overwhelmingly treat LLMs as task-automation tools and agents, the discourse around such powerful AI capabilities creates a context in which proposals to use LLMs as research participants naturally emerge. In light of these broader technological ambitions and the specific research proposals they inspire, it is both timely and necessary to critically evaluate the replacement perspective and its methodological implications, particularly to clarify the role of LLMs in behavioral and social science research more broadly, whether or not the replacement proposal gains traction.

To this end, this article examines the fundamental assumptions that underlie proposals to use LLMs as human replacements. By clarifying what LLMs actually are—computational tools that simulate linguistic behavior through statistical pattern matching and reinforcement learning (see **Box 1**)—we can better understand their appropriate role in psychology and behavioral sciences.

Six key fallacies are identified, each arising from the misinterpretation of LLMs as human replacements: the token prediction as human intelligence fallacy (conflating statistical text prediction with genuine human intelligence); the average human fallacy (assuming model outputs represent typical human responses); the alignment as explanation fallacy (interpreting similarity between model and human outputs as evidence of shared cognitive mechanisms); the anthropomorphism fallacy (attributing human-like mental states to AI); the identity essentialization fallacy (treating social categories as fixed, homogeneous traits rather than fluid, contextual identities); and the substitution fallacy (presuming LLM-generated data can directly replace human evidence without validation).

These fallacies undermine research validity—the degree to which our evidence and reasoning support the conclusions we draw (Cook & Campbell, 1979; Cronbach & Meehl, 1955; Kerschbaumer et al., 2025; Stanley & Campbell, 1963). Validity concerns in LLM research stem from both methodological choices (how we prompt and test models) and theoretical interpretations (what we claim model outputs represent). Each fallacy undermines validity in distinct ways: Some compromise our ability to make causal inferences, others limit generalizability across populations, and still others reflect misalignment between what we intend to measure and what we actually capture. The methodological and theoretical safeguards provided for each fallacy address these threats by strengthening both research design and interpretive frameworks.

Taken together, these analyses support the function of LLMs as neural language simulators. Rather than stand-ins for people, they reveal latent psychological and cognitive dynamics embedded in the text—dynamics rooted in the communication of human thoughts, attitudes, and behaviors. This approach offers both conceptual clarity and practical guidance for leveraging LLMs appropriately in behavioral and cognitive research.

**Box 1 | Language Models and the Language User Illusion**

Modern AI chatbots like ChatGPT are built on large language models (LLMs). Unlike traditional rule-based systems, LLMs process text by converting it into tokens (units of text) and mapping these to numerical vectors in a high-dimensional space (embedding). This mathematical representation enables models to capture semantic relationships between concepts—such as "king" relating to "queen" similarly to how "man" relates to "woman." The breakthrough transformer architecture (Vaswani et al., 2017) revolutionized these models through attention mechanisms that allow words to gather contextual information from surrounding text. This helps LLMs resolve ambiguities in language and handle complex linguistic tasks that once seemed to require human intelligence.

Base models (also known as "foundation" or "pretrained" models) are typically trained through autoregression—predicting each subsequent token based on preceding ones—enabling them to compress vast amounts of patterns and textual knowledge into their network weights. However, consumer-facing instruct models like ChatGPT undergo further post-training. This often involves supervised learning (fine-tuning on specific instruction–response examples) and reinforcement learning (RL; learning from feedback signals, like rewards or penalties, to improve outputs). One prominent RL technique is reinforcement learning from human feedback (RLHF), which aligns model outputs with human preferences. Beyond alignment, RL methods are increasingly used to enhance specific cognitive-like capabilities, including reasoning and tool use (e.g., interacting with web browsers or code interpreters), often through trial-and-error processes where the models learn to generate more effective or logically sound sequences of thought or action. Beyond model weights, the behavior of instruct models also depends heavily on system prompts (extensive human-authored instructions that guide model behavior), user prompts (Lin, 2024), and inference-time techniques (e.g., methodologies like retrieval-augmented generation [RAG]).

This sophisticated combination contributes to the illusion that the model possesses human-like understanding (Mitchell & Krakauer, 2023). It allows LLMs to model language in ways that go beyond traditional distributional semantics (Lake & Murphy, 2023). They capture nuanced, contextualized meanings and structures in language (Mahowald et al., 2024; Manning et al., 2020; Millière, 2024), including visual knowledge embedded in text (Jones et al., 2022; Lewis et al., 2019) and abstract concepts like space and time (Gurnee & Tegmark, 2023). When LLMs produce responses that feel empathetic or insightful, users often project human-like qualities onto them—what can be termed the "language user illusion." The fluent language generated by LLMs creates a compelling illusion of interacting with another mind, but this overlooks their fundamentally different nature from human language users. This distinction forms the critical foundation for the fallacies examined throughout this paper.

## Interpretive Fallacies in LLM-Based Research

To understand why LLMs function as simulators rather than replacements, we must first examine the conceptual errors that underlie the replacement perspective. Six fallacies emerge in discussions of LLMs as human substitutes, each revealing a different way researchers might misinterpret what LLMs are and what their outputs signify (**Table 1**).

**Table 1**

*Six Fallacies in Conceptualizing LLMs as Human Replacements*

| Fallacy | Core Misconception | Why It's Problematic | Research Implications |
|---|---|---|---|
| **Token prediction as intelligence** | Mistaking statistical pattern matching for human understanding | LLMs process language without embodied experience, consciousness, or genuine understanding | Cannot assume cognitive equivalence or generalize from LLM performance to human cognition |
| **The average human** | Assuming LLMs represent typical or average human responses | Training biases (WEIRD, temporal) and optimization goals distort LLM output representation | Cannot reliably generalize findings to diverse human populations or contexts |
| **Alignment as explanation** | Inferring shared cognitive mechanisms from similar outputs | Different architectures can produce similar results (multiple realizability); surface similarity is not mechanistic equivalence | Output alignment does not confirm shared underlying processes with humans |
| **Anthropomorphism** | Attributing human-like mental states to statistical systems | Current LLMs lack consciousness, intentions, or genuine emotional understanding | Invalidates psychological constructs requiring mental states; risks misinterpretation |
| **Identity essentialization** | Treating demographic labels as fixed, homogeneous categories | Reduces complex, fluid, and intersectional identities to stereotypes; ignores within-group diversity | Misrepresents population diversity and the construct of identity itself |
| **Substitution** | Using LLMs as primary tools for understanding human cognition or behavior | Static training data, lack of real-time adaptability, and potential for closed-loop effects | LLM data requires human validation; cannot directly replace human participants |

### Token Prediction as Human Intelligence Fallacy

LLMs accomplish tasks once reserved for humans through a layered process: They are initially pretrained to predict sequential tokens; then post-trained via fine-tuning and reinforcement learning; and finally, their reasoning behavior is shaped by guidance from well-crafted prompts and inference-time techniques (**Box 1**). This process equips LLMs with intricate linguistic knowledge (e.g., syntactic rules, semantic relations)—often termed *formal* linguistic competence (Mahowald et al., 2024). Crucially, these models do more than memorize patterns; they demonstrate the ability to navigate complex, context-dependent tasks in both language processing and generation (Mahowald et al., 2024; Millière & Buckner, 2024a). Through learning and contextual adaptation, LLMs can infer underlying task structures and generate appropriate responses across varied domains—a form of *instrumental* knowledge that supports problem-solving in diverse settings (Yildirim & Paul, 2024).

Yet, at their core, LLMs have no minds (Searle, 1980) but are autoregressive statistical models that manipulate language—a task different from human cognition, producing a kind of ungrounded intelligence. They manipulate language by leveraging statistical patterns learned during training; while these patterns encode aspects of *derivative* meaning present in text-corpus relationships (e.g., "Paris"–"France," "fire"–"hot"), the models themselves are yet to access this meaning through direct, grounded experience (e.g., visual qualia, proprioception). This mirrors Church encoding in lambda calculus, where data and operations are expressed purely through abstract functions, defined by how they transform inputs to outputs, rather than by inherent, grounded meaning (Church, 1936). This is in contrast to human intelligence, which integrates general and specialized capabilities through evolutionary adaptations and developmental learning (**Box 2**)—and is "grounded in one's embodied physical and emotional experiences" and "deeply reliant on one's social and cultural environments" (Mitchell, 2024).

A fallacy emerges when we mistake the capacity of LLMs to predict language patterns and solve cognitive problems for genuine human-like understanding or intelligence. To unpack this fallacy, it is helpful to distinguish between two levels of understanding: functional understanding (reliably producing appropriate outputs, as when a dog successfully catches a ball) and reflective understanding (consciously grasping underlying principles, as when humans can articulate why the ball follows a particular trajectory). While LLMs may demonstrate considerable functional understanding by generating contextually appropriate responses, they currently lack the reflective understanding that emerges from embodied experience and consciousness. Consequently, mistaking their output for genuine cognition would misrepresent what is being measured; the model's linguistic performance, based on statistical token prediction, risks being conflated with human cognitive processes that integrate both functional and reflective understanding. The "intelligence" of LLMs is thus operationally different from the human cognition it might appear to simulate.

This fallacy manifests in two major ways. First, while LLMs process vast amounts of text reflecting real-world information and human experience, their tokens operate without direct, first-hand sensorimotor contact with the world. Consequently, they lack embodied referents and lived experience derived from such interaction (Leivada et al., 2023). This critical absence of grounding is demonstrated by how text-based LLMs, while capable of representing non-sensorimotor features of human concepts (e.g., emotional valence), systematically fail to capture sensorimotor features, particularly those related to motor actions. Indeed, the alignment between LLM and human conceptual representations diminishes markedly from non-sensorimotor to sensorimotor domains (Xu et al., 2025).

Second, this absence of first-person, sensorimotor grounding limits the models' connection to the physical and social realities they aim to simulate. This hinders the development of fully *functional* linguistic competence—the use of language to achieve goals in the world (Mahowald et al., 2024). It also restricts their capacity to acquire the kind of *worldly* knowledge needed for robust world models—internal representations that are structure-preserving and behaviorally efficacious in real-world interactions (Webb et al., 2023; Yildirim & Paul, 2024). Such models, including cognitive maps, body schemas, and spatial schemas, support embodied reasoning and action, which remain beyond the scope of current models (Wicke & Wachowiak, 2024).

A critical consequence of this fallacy is that when researchers attribute changes in model outputs to experimental manipulations without recognizing the underlying statistical nature, they inevitably risk misinterpretation. Consider a simple prompt: "*Mao Zedong was…*" Unlike interacting with another mind, when we engage with a chatbot, we are not seeking its opinion—despite the compelling illusion thereof—but rather making a *computational* request: Given the statistical distributions in the language model, what sequence is most likely to follow these words? Models trained for neutrality will likely provide correspondingly factual responses. Indeed, the character of such responses is shaped by the model's training, geopolitical alignment, and prompt language. For example, DeepSeek-R1—pretrained mostly on English and Chinese text, and subsequently aligned with government regulatory directives (**Box 1**)—showed substantially higher proportions of Chinese-state propaganda and anti-U.S. bias compared with ChatGPT o3-mini-high, which was pretrained primarily on English text. This bias was most pronounced with Simplified Chinese queries, diminished with Traditional Chinese inputs, and was nearly absent when queried in English (Huang et al., 2025).

Fundamentally then, the model, unlike a human, has no communicative intent, no opinion of, attitude toward, or belief about Mao, and no intrinsic capacity to tell the truth—it just models a distribution of token sequences based on the training texts (Shanahan, 2024). While they manifest proficient language *use* that goes beyond simply retrieving pre-recorded text strings, they are not true language *users* in the philosophical sense. They do not possess intrinsic meaning, communicative intentions, or other internal states essential to human language users (Block, 1981).

Indeed, model outputs can be highly sensitive to seemingly trivial prompt variations—variations that human language users would tolerate (Ivanova, 2025). Small changes in prompt language, wording, order, or context can produce drastically different outputs not because of the manipulated variable of interest, but due to the model's inherent sensitivity patterns—a phenomenon that contrasts with how humans process language.

Additionally, findings from LLM studies cannot be reliably generalized to human cognition. The gap between token prediction mechanisms and human cognitive processes fundamentally limits what we can infer about real human populations from model behavior, even when performance patterns may appear superficially similar. As philosopher John Searle illustrated in his "Chinese Room" thought experiment, a system might process symbols according to rules without understanding their meaning—just as a person following instructions to manipulate Chinese characters could produce appropriate responses without comprehending Chinese. When models generate coherent language without the intentionality, consciousness, and direct grounding in real-world experience that characterize human intelligence (Searle, 1980), they are capturing linguistic *form* (the observable structure of language) rather than true *meaning* (the communicative intent behind language) (Bender & Koller, 2020). Therefore, to confuse cognitive algorithms for cognition, or models of the mind for the mind itself, represents a category mistake, both ontologically (misidentifying their nature) and epistemologically (misunderstanding their knowledge).

No doubt LLMs (and AI in general) will continue to advance. But, as detailed in **Box 3**, even these improvements face fundamental conceptual limitations regarding embodied experience and grounding. Knowledge, as constructivists like Jean Piaget have argued, is built from sensory experiences and perceptions, which are then layered with symbols and categories over time. Without such experiences, LLMs are like an artificial version of the proverbial Mary studying everything about color in a black-and-white room her entire life (Jackson, 1982). One can argue that when Mary leaves her room and sees color for the first time, she learns something new—what it is like to see something pink (Jackson, 1986). If so, a great deal more is at stake for language models, which experience neither color nor anything. Indeed, unlike Mary, who can use her other experiences as a scaffold for understanding color—much like Helen Keller using associations from senses like touch and smell to construct a color scheme despite being blind and deaf ("*Pink makes me think of a baby's cheek, or a gentle southern breeze*")—LLMs have none of this kind of sensory scaffolding.

Essential aspects of human cognition—emotions, intuition, and other subjective experiences—so far remain incomputable and therefore cannot be fully replicated by computational algorithms. Even if everything about human cognition were theoretically computable, actually capturing the full depth and complexity of human thought processes would likely exceed computational feasibility. The interplay of perception, memory, emotions, and social contexts creates patterns so intricate that complete replication would require exponentially increasing computational resources. In other words, creating AI systems that genuinely replicate human cognition—systems behaving like humans under all circumstances—may be fundamentally intractable from a computational perspective (Van Rooij et al., 2024). AI systems—constrained by their current algorithms, data, and embedded assumptions—face inherent challenges in replicating human cognition. This limitation parallels Gödel's famous discovery in mathematics: Any formal system sophisticated enough for basic arithmetic will contain true statements that cannot be proven within that system (Fokas, 2023). Similarly, AI systems face intrinsic constraints in capturing all aspects of human thought within their computational frameworks.

To mitigate these issues, consider the following.

1. **Model selection and settings**: Different model architectures may capture different aspects of linguistic behavior (Dettki et al., 2025). Choose models based on their specific capabilities for the research question rather than assuming larger or more advanced models automatically approximate human cognition better (Zan et al., 2025; Zhou et al., 2024). Compare base models with fine-tuned versions to understand how optimization objectives affect outputs relative to human responses (Binz & Schulz, 2023a; Yax et al., 2024). Evaluate the impact of model parameters such as temperature (Li & Qi, 2025).
2. **Prompt design**: Test model sensitivity to prompt variations to distinguish between robust patterns and artifacts of specific phrasings (Brucks & Toubia, 2025). This helps identify when models are exhibiting systematic "reasoning" versus merely responding to surface-level cues in ways humans would not (Ivanova, 2025).
3. **Interpretations and applications**: Validate LLM outputs against human data, particularly when studying cognitive processes, to establish appropriate boundaries for generalizing from LLM experiments to human psychology. Make explicit that LLM performance, however impressive, stems from statistical prediction rather than human-like understanding (Shiffrin & Mitchell, 2023).
4. **Ethics**: Acknowledge model limitations in research reporting, particularly the inherent distinction between token prediction and human cognition, to prevent misinterpretation and reduce the risk of anthropomorphizing LLMs in scientific literature (Ibrahim & Cheng, 2025).

**Box 2 | Biological versus Machine Optimization**

Natural selection introduces a different optimization mechanism from machine training, which directly impacts how we should interpret intelligence in both systems. Biological pressures for survival and reproduction embed implicit assumptions and constraints within neural systems—termed ecological inductive biases (Richards et al., 2019). These biases enable humans to learn efficiently from physical and social interactions, manage limited cognitive resources by prioritizing selective inputs and tasks, and adapt to uncertainty using shortcuts and flexibility. Such evolved constraints stand in contrast to the engineered bias of language models, which arise from their objective function (e.g., next-token prediction, reward maximization), design (e.g., embeddings, transformer architecture), and training data composition. Other computational models similarly have their distinctive inductive biases; for example, convolutional neural networks (CNNs) learn patterns by focusing on local spatial relationships, as their convolutional filters operate over small image regions.

A crucial distinction is that biological evolution carries historical baggage that constrains optimization in ways that shape cognitive capabilities. When environments suddenly change, biological systems cannot rebuild from scratch but must adapt using existing neural machinery, often creating suboptimal yet functional solutions. This evolutionary path dependency is further constrained by the DNA bottleneck, limiting intergenerational information transfer: All adaptive changes must be encoded in the genome, which can only store a fraction of an organism's lifetime learning. In contrast, AI systems face no such constraints. They can be redesigned from the ground up with each iteration, and their "knowledge" can be transferred directly via perfect copying, distillation into smaller models, or selective merging of capabilities—all without the genetic limitations that restrict biological information transfer.

These disparities between language models and humans in their optimization pressures manifest in both internal architectures and external behaviors (Doerig et al., 2023). Key architectural differences include:

- Computation mechanisms: Models rely on digital computation and continuous activation functions, contrasting with the analog computation and all-or-none neuronal firing characteristic of the human brain;
- Information flow: Models predominantly use feed-forward processing, whereas the human brain exhibits extensive feedback modulation across interconnected regions; and
- Learning algorithms: Models depend on backpropagation—an algorithm that adjusts network weights based on output errors—for learning, which is biologically implausible due to its reliance on explicit error signals, symmetric weight updates during forward and backward passes, global error propagation throughout the entire network, and discrete, stepwise updates after processing batches of data.

These architectural differences produce observable behavioral distinctions between LLMs and human cognition:

- Data efficiency: LLMs exhibit low data efficiency, as evidenced by sublinear scaling laws—while increasing data and model size leads to improved performance, achieving incremental improvements requires exponentially more resources;
- Energy consumption: LLMs display low energy efficiency, consuming megawatts of power compared to the ~20 watts used by the human brain;
- Memory stability: Models show susceptibility to catastrophic forgetting, where learning new tasks causes the model to forget previously learned tasks;
- Learning approach: Current LLMs lack active exploration and continual learning—they do not retain prior interactions or instructions across sessions—unlike humans, who learn cumulatively over time; and

- Input robustness: Models demonstrate fragility to minor input perturbations and noise, exemplified by the sensitivity of LLMs to prompt variations, namely, brittleness.

These differences in optimization mechanisms, architectural constraints, and resulting behaviors illustrate why token prediction, however sophisticated, produces a qualitatively different form of intelligence than human cognition—and thus LLMs cannot simply replace human participants in psychological research.

**Box 3 | Technical versus Conceptual Limitations in Model Simulation of Human Behavior and Cognition**

When evaluating language models as simulation tools for psychological science, distinguishing between technical and conceptual limitations clarifies both current research constraints and future prospects. Technical limitations arise from current implementation practices and can be mitigated through methodological refinements. These include: (1) training data biases—predominantly WEIRD-skewed text corpora with specific temporal cutoffs; (2) algorithmic artifacts—such as those introduced by RLHF that systematically alter response patterns; (3) architectural constraints—including limited context windows and memory capabilities; and (4) prompt sensitivity—where minor wording variations produce substantially different outputs. These limitations affect simulation fidelity and require various methodological controls, but are potentially addressable through engineering advancements within the existing paradigm.

Progress in addressing technical limitations is already underway. Multimodal LLMs (MLLMs) now process information across text, images, audio, and video. Reasoning-focused models (sometimes referred to as large reasoning models) integrate reinforcement learning with chains of thought, explicitly narrating intermediate reasoning steps. This scaffolding promotes deeper analysis and improves performance on tasks requiring structured, multi-step inference across domains like science, math, and code. As models learn from increasingly diverse digital media, they develop more sophisticated representations of the physical world, potentially capturing dynamics beyond human sensory reach (e.g., at micro and macro scales) and improving aspects of physical common sense derived from text alone.

Conceptual limitations, by contrast, reflect intrinsic discrepancies between model design and human cognition that define the boundaries of the current paradigm. These include: (1) grounding—LLMs lack embodied experience that situates cognition in physical and social contexts, operating solely through abstract symbol manipulation; (2) agency—LLMs have no intrinsic goals, desires, or motivations driving behavior, functioning instead as sophisticated prediction systems; and (3) subjectivity—LLMs lack conscious experience that introduces qualitative dimensions to human cognition, including emotional valence and phenomenal awareness. These limitations constrain the construct validity of LLMs as psychological models in ways not readily addressed through incremental engineering.

Even rich observational learning from video may not equate to the embodied knowledge humans acquire through direct, multisensory interaction. Without lived experience, models might still struggle to fully replicate crucial aspects of human intelligence, such as deeply grounded physical intuition, nuanced social common sense, and certain forms of mathematical reasoning (S. Lee et al., 2024; McCoy et al., 2024; Mirzadeh et al., 2024), even when they mimic or surpass human performance in other ways. For example, larger models improve in certain domains (e.g., some challenging pattern recognition tasks) but still fail at seemingly simple tasks that humans would expect them to handle easily (Zhou et al., 2024). The depth of understanding gained without interactive, multisensory embodiment may differ from human cognition (Jones & Bergen, 2024), resembling, to some extent, the difference between inferring object properties from descriptions or observations versus learning through direct manipulation and sensory feedback, where crucial experiential nuances can be missed (cf. Kim et al., 2019, regarding visual properties and congenital blindness).

Paradoxically, solutions to these conceptual limitations may render future AI systems *less* analogous to human cognition even while they gain superhuman abilities. For instance, addressing grounding limitations through embodied robotics with RL would introduce fundamentally different learning processes from both current text-based training and human development—such as active, continuous learning from direct experience via grounded rewards from the world using non-human tools, rather than relying on human input. Similarly, systems designed to overcome reasoning limitations might develop internal representations and goals with no human analogue. For example, as relentless problem-solving agents, these systems may exhibit a distribution of competence that differs from that of an average person.

As the field progresses beyond current paradigms, the relationship between machine and human cognition will continue to evolve, requiring ongoing reassessment of the role of LLMs in psychological science.

**The Average Human Fallacy**

Mischaracterizing the nature of intelligence in LLMs may not matter as much if they behave or perform like the average human, enabling them to functionally replace human participants. But this assumption commits the average human fallacy, which misrepresents what LLM outputs signify.

Consider the engineering purpose and approach of LLMs. They are explicitly developed to outperform humans, as measured by a broad range of benchmarks and standardized tests—an engineering feat bolstered by continually improved design and algorithms, and access to more data and compute, while free from biological limitations (see also **Box 2**). In actual tests, for example, GPT-4 outperforms humans in detecting and interpreting irony, recognizing indirect requests or hints in conversation (Strachan et al., 2024), analogical reasoning tasks (Webb et al., 2023), and probabilistic reasoning tasks like the Linda/Bill problems and the bat-and-ball problem (Yax et al., 2024), but underperforms humans in tasks like faux pas tests (Strachan et al., 2024). This contrasts with the assumption of the replacement view that LLM responses mirror average human judgments from the training data (Dillion et al., 2023) or the majority's mainstream opinions (Qu et al., 2024).

The generalizability challenge is multifaceted. First, LLMs are not representative samples of any defined human population, let alone an "average" human. Their training data (mostly online text) exhibit systematic biases—predominantly Western, Educated, Industrialized, Rich, and Democratic (WEIRD) populations, especially those who are hegemonic, young, and publicly expressive (Crockett & Messeri, 2023; Santurkar et al., 2023; Tao et al., 2024). This biased training creates a complex landscape for psychological simulation: While LLMs might adequately capture behaviors that are largely universal across human cultures (certain basic cognitive or emotional phenomena, perhaps), they face limitations when simulating those known to vary across cultures and groups—from number representations to personality traits and moral reasoning. Further complicating this picture is our incomplete understanding of which psychological phenomena are universal versus culturally variable. This uncertainty means researchers must exercise particular caution when using WEIRD-biased LLMs—or any LLMs for that matter—to simulate potentially culture-dependent phenomena.

In tandem with spatial biases, the temporal distribution of the data is more concentrated in recent history, with the model's understanding of the past filtered through the lens of contemporary languages and norms (Ziems et al., 2024). This presentist and recency bias risks temporal flattening, wherein historical and contemporary voices are homogenized, obscuring the richness of historical diversity in human thought. This is problematic for probing thoughts and behaviors that have evolved over extended periods—such as the way people think about concepts like gender, race, and class (Kozlowski & Evans, 2024). The models may lack the contextual richness and temporal granularity needed to understand how these concepts were discussed in different periods, potentially reinforcing contemporary biases when trying to understand people from earlier time periods.

Beyond training data biases, the engineering goals of LLMs further compromise their representational accuracy. The goal of developing advanced LLMs—to provide accurate, helpful answers to users—contrasts with the study of human cognition, which is replete with inaccuracies, biases, shortcuts, and idiosyncrasies. For instance, in language comprehension, people often settle for a partial and sometimes inaccurate understanding that is nevertheless sufficient for the task at hand—"good-enough representations" (Ferreira et al., 2002). In decision making, when emotions are induced in bargaining games and repeated cooperation games, GPT-4 tends to maintain consistent, rational decision-making, in contrast to humans (Mozikov et al., 2024).

Due to model limitations, opacity, and the common use of RLHF, such models cannot be assumed to represent the "average" of their training data. Responses may reflect hallucination, the influence of RLHF, or other bias reduction efforts; they could also simply regurgitate specific instances, examples, or strategies from their training data, particularly when the query is well represented in the data (Aher et al., 2023; Binz & Schulz, 2023a; Shiffrin & Mitchell, 2023). For example, when LLM responses are altered through RLHF, they can deviate from the original training data—and may even exacerbate their misalignments with non-dominant views (Santurkar et al., 2023). In this process, the communicative intent of developers (and data labelers) shapes model outputs to prioritize goals like accuracy and helpfulness, rather than reflecting "raw" human-like responses from the training texts.

This raises concerns about whether the outputs genuinely represent human responses. If the psychological construct being measured is meant to be "average human response," the LLM fails to operationalize this construct due to the representational biases described above. The models often become "too neutral, detached, and nonjudgmental," lacking selfhood and initiative (Ye et al., 2024), and exhibiting homogeneous personality profiles—high in agreeableness and low in neuroticism (Li & Qi, 2025; Pellert et al., 2024). While this fine-tuning process can reduce certain biases (Hu et al., 2025), enhance accuracy, and make the model more pleasant to interact with, it also weakens its ability to reflect the actual attitudes and thoughts present in human texts (Harding et al., 2024) and to produce diverse responses (Murthy et al., 2024). Such alterations may also introduce new preferences or biases from the feedback, including sycophancy (a tendency to generate outputs that are excessively user-pleasing or conform to perceived desirable responses), making it challenging to rely on RLHF-tuned LLMs as accurate indicators of human thought and judgment (Park et al., 2024). In survey responses, for example, LLMs do not share human response biases, with more pronounced discrepancies in RLHF-tuned models (Tjuatja et al., 2024). Likewise, within the GPT-3 family, fine-tuned models exhibited higher propensity for conjunction fallacy and intuitive reasoning relative to base models (Yax et al., 2024).

The practical implications of these misalignments are evident in actual response patterns. LLM responses have been found to mischaracterize marginalized groups, as evidenced by out-group imitation rather than in-group description (Wang et al., 2025), and misrepresent sampled groups, as demonstrated by upward bias in mean ratings of the Big Five personality traits (Niszczota et al., 2025), as well as in other surveys and tests (Hagendorff et al., 2023; Sarstedt et al., 2024; Tjuatja et al., 2024). In addition to these shifts in average responses, LLMs also fail to capture the nuances and heterogeneity of human responses, producing flattened, oversimplified portrayals of various groups (Wang et al., 2025).

While diversifying the training corpus to include more languages and cultural contexts helps to broaden representations, achieving global representations is ultimately a long-term challenge (Lin & Li, 2023). In the foreseeable future, the quantity and quality of available non-English training texts will remain impoverished compared to English as the lingua franca.

To mitigate the average human fallacy, key considerations include explicitly defining represented populations, potentially using specialized models, testing prompts and models for consistency and bias, validating responses against human data, and transparently reporting representational limitations. Specifically:

1. **Model selection and customization**: Be explicit about which populations are represented in the training data and how this affects generalizability claims. Consider using specialized models fine-tuned on representative human data and, importantly, verify performance with actual samples from those populations (Gao et al., 2024; Suh et al., 2025).
2. **Prompt design**: Test models across multiple prompting styles to assess the consistency of simulated responses and determine sensitivity to superficial variations (Li & Qi, 2025; Momennejad et al., 2023; Tao et al., 2024). Compare outputs from different LLM versions and architectures to identify systematic biases in representation.
3. **Interpretations and applications**: Validate LLM responses against human samples, particularly for claims about specific populations or demographic groups, and report cases where LLM responses diverge from human patterns. Avoid generalizing beyond domains where empirical validation has been established (Abdurahman et al., 2025).
4. **Ethics**: Document potential biases and acknowledge limitations in representativeness when reporting results from LLM simulations (Abdurahman et al., 2025).

**Alignment as Explanation Fallacy**

High output alignment between LLM responses and human data has been suggested as evidence that LLMs can sometimes replace human participants (Dillion et al., 2023). While such alignment suggests that LLMs might capture something mechanistic about human behavior, it is fallacious to assume mechanistic equivalence—that the model explains human cognition (Guest & Martin, 2023) or can replace human participants for theoretical understanding (see **Box 4**). Treating such alignment as sufficient for mechanistic equivalence constitutes the alignment-as-explanation fallacy. This fallacy arises when output similarity is conflated with representational and processing similarity—that is, when similar outputs are taken as evidence that LLMs engage the same psychological constructs or cognitive mechanisms as humans, overlooking the differences between the statistical pattern-matching of current LLMs and the embodied, contextual cognitive processes of humans.

Indeed, a fundamental problem of applying human-centric tests and concepts to LLMs—such as theory of mind or emotional understanding—is the assumption that they engage with information in ways similar to humans, presupposing background psychological mechanisms that may be absent or irrelevant for LLMs (**Box 1**). These anthropomorphic assumptions undermine the construct validity (i.e., the degree to which a test measures the specific psychological construct it purports to measure) of psychological tests in LLMs (Millière & Buckner, 2024b). For example, response confidence in LLMs, as measured by token probability, may differ from human self-reports. Similarly, emotional intelligence—the ability to perceive, understand, manage, and use emotions in oneself and others—encompasses self-awareness, empathy, emotional regulation, and social skills, all rooted in subjective experiences that are absent in LLMs. Model "understanding" of emotions is purely expressive and syntactic, driven by learned associations between words and concepts rather than experiential, visceral insight or an internalized understanding of mental states. So, when LLMs perform like humans on tests of emotional understanding—in either overall score or response pattern (Wang et al., 2023)—this alignment is at the surface level, rather than reflecting genuine equivalence.

Apparent alignments between LLM and human responses may also be artifacts of specific task formulations rather than evidence of robust, human-like reasoning capabilities. One clue for a lack of true competence is a dissociation between accuracy and the model's explanations of its responses—namely, reasoning (Leivada et al., 2023). Another critical test is prompt sensitivity, examining how performance is affected by superficial alterations to the prompts—changes to which humans typically show little to no sensitivity or respond to differently. Such prompt sensitivity has been documented in LLMs in survey responses—RLHF-tuned models can be highly sensitive to changes like typos (Tjuatja et al., 2024)—and in tasks like reasoning (Binz & Schulz, 2023a; Dasgupta et al., 2022; Yax et al., 2024), decision making (Binz & Schulz, 2023b; Suri et al., 2024), theory of mind (Strachan et al., 2024), moral judgments (Oh & Demberg, 2025), and more (Kamoi et al., 2024; McCoy et al., 2024).

For example, LLMs but not humans improved their performance in reasoning tasks when the instruction included the phrase "let's think step by step" (Yax et al., 2024), a form of chain-of-thought prompting. Conversely, while LLMs performed at ceiling in a false belief test just like human participants, they struggled when small changes were made to the formulation of false belief scenarios—suggesting syntactic pattern processing rather than robust reasoning (Strachan et al., 2024). These sensitivities to minor variations highlight how apparent alignment can mask differences in underlying processes, creating a threat to internal validity—the ability to draw firm conclusions about causal relationships—when the observed performance is misattributed to the intended manipulation rather than to superficial linguistic patterns.

But even when performance reflects underlying abilities, it is not clear whether models employ mechanisms similar to those of humans. Given that different systems can achieve the same outcome through different mechanisms—known as multiple realizability (Bowers et al., 2023; Guest & Martin, 2023), as in telling time with digital versus mechanical clocks—it is unwarranted to assume mechanistic or even functional equivalence (**Box 4**). Indeed, it is notoriously difficult to understand exactly what LLMs have learned. Beyond inherent differences between machine and biological intelligence in their architecture and algorithms, LLMs are trained on datasets much larger than what human learners experience. Differences in mechanisms may manifest as distinct response characteristics, including (1) context sensitivity, such as prompt sensitivity or performance variations across different vignettes (Yax et al., 2024); (2) response patterns, such as variability in open-ended responses (Li et al., 2024), item-by-item performance variability (Wang et al., 2023), and correlation of accuracy with confidence (Yax et al., 2024); and (3) error types and consistency, such as errors arising from cognitive demands versus those from item wording or familiarity (Yax et al., 2024).

This raises a fundamental question: When LLMs perform at human levels in psychological tasks, what does it tell us about the capacity of these models? Performance can reflect true competence—some underlying abilities—or something more superficial, like pattern memorization (Gao et al., 2024) or other surface-level cues or pure chance. Conversely, underperformance can reflect something other than incompetence, such as processing limitations or ineffective prompting (Firestone, 2020). Thus, to establish true competence, model outputs should be sensitive to changes in the task-relevant inputs but insensitive to irrelevant changes (Harding & Sharadin, in press).

To mitigate the alignment-as-explanation fallacy, safeguards include testing alignment robustness against perturbations, examining reasoning processes, using causal interventions, performing cross-domain validation, and explicitly acknowledging the limits of inferring shared mechanisms. Specifically:

1. **Model testing and validation**: Test whether the alignment between LLM and human responses is robust to perturbations in task structure, prompt wording, and contextual variations that shouldn't affect performance (Oh & Demberg, 2025). If the performance varies significantly with superficial changes, this suggests a lack of true construct equivalence (Firestone, 2020).
2. **Process tracing**: Use methods such as chain-of-thought prompting to examine the reasoning paths that LLMs use to arrive at answers, comparing these with human reasoning protocols (Bao et al., 2024). Significant differences in reasoning processes, even when outputs align, would suggest different underlying mechanisms.
3. **Causal interventions**: Implement interventions that test specific hypotheses about psychological mechanisms. For example, if humans and LLMs appear to use similar heuristics, test whether manipulations designed to affect those heuristics produce comparable effects (GX-Chen et al., 2025). Examine the causal effects of training data on model output through fine-tuning (Hu et al., 2025).
4. **Cross-domain validation**: Examine whether alignment in one domain generalizes to related domains that should engage similar cognitive processes. Limited transfer may suggest different underlying mechanisms despite surface alignment (Zan et al., 2025). The same applies to validation across different languages of the prompt, such as English and Chinese (Jin et al., 2024).
5. **Explicit limitation acknowledgment**: When reporting alignments between LLM and human responses, acknowledge the limitations of inferring shared mechanisms, noting their distinct architecture and learning history (Abdurahman et al., 2025).

**Box 4 | Functional versus Mechanistic Equivalence**

Functional equivalence occurs when two systems produce the same outputs given the same inputs under specified conditions—like how both a digital watch and a sundial tell accurate time on a clear day, despite different mechanisms. Mechanistic equivalence, in contrast, requires shared internal causal processes—addressing *how* results are produced, not merely *that* they are produced. Because distinct mechanisms can yield identical functions (multiple realizability), functional alignment alone cannot establish mechanistic alignment.

In psychological research, an LLM that correctly answers false belief questions demonstrates functional alignment with humans, but evidence from prompt sensitivity studies reveals mechanistic divergence. Even "perfect" behavioral correlation therefore licenses, at most, a claim of functional parity. Demonstrating mechanistic parity would demand convergent evidence from: (1) input perturbation tests—performance should degrade in the same pattern as humans when task-relevant cues are modified; (2) internal-state homology—representational similarity analyses should reveal shared information coding; and (3) causal intervention—disabling a model component should mirror the effect of corresponding cognitive impairments in humans (Firestone, 2020; Guest & Martin, 2023).

Differentiating functional and mechanistic equivalence has implications for three levels of LLM–human similarity:

- **Level 1: Pragmatic simulation.** If the research goal is purely functional—pre-testing vignettes, forecasting average responses, relationship counseling—then output equivalence between LLMs and humans may suffice regardless of underlying mechanisms. Here, the simulation perspective is interested primarily in *what* is produced, not *how* it is produced. For example, if an LLM reliably predicts which survey items show ceiling effects, its utility as a rapid prototyping tool remains valid even if its internal processes differ from human cognition.
- **Level 2: Explanatory inference.** If the research goal is theoretical—testing models of cognition, emotion, or moral reasoning—then superficial behavioral alignment becomes insufficient. Internal and construct validity demand evidence that similar underlying mechanisms are at play. An LLM might generate human-like responses to reasoning tasks without engaging in human-like reasoning—just as a calculator produces accurate math answers through processes entirely unlike human calculation.
- **Level 3: Phenomenological attribution.** Even if perfect functional and mechanistic overlap were demonstrated, this would not automatically confer phenomenal states—subjective experiences like visceral, emotional empathy. Claims that LLMs "have beliefs," "feel empathy," or "experience understanding" involve a third level of equivalence that requires additional, presently unknown properties (e.g., global workspace dynamics, biological embodiment, or other factors that might give rise to consciousness). Consequently, attributions of subjective experience to LLMs remain largely speculative until such criteria are articulated and empirically evaluated.

Thus, LLMs can serve as valuable functional simulators for many research purposes without needing to process and experience the world as humans do. However, researchers must be transparent about which level of equivalence they are claiming and provide appropriate evidence for that level. The danger lies not in using LLMs as Level 1 simulators, but in making unwarranted leaps to Level 2 theoretical inferences or Level 3 phenomenological attributions—shortcuts that threaten validity.

**Anthropomorphism Fallacy**

Beyond misalignments due to algorithms, purposes, and implementations, another issue with the replacement view is that it leads to anthropomorphism of LLMs (Crockett & Messeri, 2023; Shiffrin & Mitchell, 2023). Their fluent, human-seeming responses can trigger an enhanced ELIZA effect: the tendency to attribute human-like understanding to systems that merely manipulate symbols, creating a compelling "language user" illusion (**Box 1**). Indeed, when researchers refer to "the minds of language models" or "the machine minds of LLMs" (Dillion et al., 2023), such terminology—whether used metaphorically or literally—can inadvertently encourage teleological bias, attributing purposes and goals to LLMs.

The anthropomorphism fallacy misinterprets LLM responses, treating statistical artifacts as expressions of an inner mental life. When researchers ascribe human-like mental states—such as beliefs, intentions, or consciousness—to LLMs based on their linguistic output, they risk invalidating measures of psychological constructs. Indeed, many such constructs, including attitudes and emotions, presuppose a mental architecture that current LLMs lack.

In many situations, using anthropomorphic language in conversation is natural—even useful. However, with AI like LLMs—which currently lack documented markers of consciousness yet produce remarkably coherent conversations (Shardlow & Przybyla, 2024)—this semblance seduces users to interpret model behavior through the lens of folk psychology, attributing "beliefs" or "consciousness" to these systems (Colombatto & Fleming, 2024). Indeed, the chatbot service provider Character AI invites users to meet AIs that "feel alive."

While the philosophical debate about machine consciousness remains open (Butlin et al., 2023), ascribing emotions or intentions to current LLMs (saying they "believe" or "think") risks creating impressions that outpace their demonstrated capabilities. As of mid-2025, no publicly available LLM exhibits clear markers of phenomenal awareness or intentional agency. Such attribution gaps can lead researchers and the public to either overestimate or underestimate these systems' capabilities (Crockett & Messeri, 2023; Shanahan, 2024), potentially shaping AI policy in ways disconnected from technological reality (Lin, 2025). Perhaps most concerning, this conflation may dilute our understanding of distinctly human qualities—feelings, thoughts, and virtues—thereby diminishing their meaning (Vallor, 2024).

The fallacy also leads researchers to draw invalid inferences from LLM outputs. When interpreting model responses as expressing human-like mental states rather than statistical probabilities, they may fail to account for the statistical artifacts inherent in token prediction systems. Similarly, findings may be inappropriately generalized from anthropomorphized LLMs to human populations without recognizing the essential difference between statistical text generation and human psychological processes.

The replacement perspective thus risks anthropomorphizing algorithms and mischaracterizing their nature—a conceptual error that invites misunderstandings and misinterpretations. For example, one such mischief is that "any given LLM can act as only a single participant" (Dillion et al., 2023). Yet unlike humans, who are influenced by a unique combination of personal experiences, emotions, and cognitive biases, LLMs are not limited to a single perspective but generate responses based on their vast, diverse dataset. This means that, depending on the prompt and context, the same LLM can produce a range of patchwork responses, each reflecting different viewpoints or types of reasoning (Santurkar et al., 2023). This variability is not indicative of a singular, consistent "mind," but rather of a multifaceted tool capable of simulating diverse perspectives. A teenager, a senior citizen, a subject matter expert, or a layperson: LLMs can role-play various characters or personas (Shanahan et al., 2023). This chameleon-like ability highlights LLMs as tools for linguistic simulation, not as human participants.

Mitigating the anthropomorphism fallacy requires using precise language, maintaining conceptual clarity about LLMs' lack of mental states, documenting technical settings, applying simulation-based interpretive frameworks, and specific researcher training. Consider:

1. **Language and framing**: Use precise, non-anthropomorphic terminology when describing LLM outputs. Instead of saying an LLM "believes" or "feels," opt for terms like "produces," "generates," or "outputs" to accurately reflect their statistical nature (Ibrahim & Cheng, 2025; Shanahan, 2024).

2. **Conceptual clarity**: Explicitly acknowledge in research designs and reports that current LLMs do not possess mental states or consciousness (Shardlow & Przybyla, 2024). Define psychological constructs carefully, noting when they inherently depend on mental states that LLMs lack.
3. **Documentation practices**: Document the specific LLM, version, prompt design, and parameter settings used to generate responses (Lin, 2025), emphasizing the technical rather than psychological aspects of the process.
4. **Interpretive frameworks**: Develop and apply interpretive frameworks that treat LLM outputs as simulations rather than expressions of beliefs or attitudes (Ibrahim & Cheng, 2025). This includes distinguishing between "simulated beliefs" and actual beliefs when reporting results.
5. **Education and training**: Provide training to research teams on the mechanisms of LLMs and the risks of anthropomorphic interpretations (Lin, 2025). Foster a research culture that maintains conceptual precision when discussing AI capabilities.

### Identity Essentialization Fallacy

Under the replacement perspective, prompting often invokes identity labeling, such as instructing the LLM to act as or adopt the identity of "White man," "Black woman," "Chinese," or "American"—as if such labels describe innate, static, homogeneous social groups, each entailing a specific set of behaviors (Chuang et al., 2024; Wang et al., 2025). This approach constitutes an identity essentialization fallacy that caricatures how identity operates in human populations.

When simple demographic labels are used to prompt LLMs, there is often an implicit assumption that these can generate responses representative of real human populations. Such prompting approaches treat social categories as static and homogeneous, ignoring the vast diversity within any demographic group. When LLMs generate responses based on these simplified identity prompts, they may produce stereotyped or inaccurate outputs that fail to capture the nuanced realities of actual human populations (Li & Qi, 2025), limiting what we can learn about real-world contexts and populations (Lahoti et al., 2023; M. H. Lee et al., 2024). Indeed, demographic prompting can even reduce alignment with human judgments (Sun et al., 2025).

In colloquial exchanges, essentialist language about social categories—from "artists are eccentric" to "women are nurturing"—is convenient and also meaningful. But in empirical research, identity essentialization masks the fluidity and diversity inherent within any demographic, overlooking individual nuances and intersectionality while reinforcing stereotypes and biases prevalent within society, thus overestimating group differences (Namboodiripad et al., 2023; Prentice & Miller, 2006).

Identity is not a static, unitary construct that can be captured by a single demographic label—it is fluid, contextual, and intersectional. When researchers use essentialist prompting techniques, they misrepresent the psychological construct of identity itself, reducing rich, complex human experiences to one-dimensional categories. This reductive operationalization fails to capture how various aspects of identity interact, how identity salience shifts across contexts, and how individuals negotiate multiple, sometimes contradictory, identity facets.

This is not to deny the importance of identities, nor to advocate for identity-blindness. As pervasive societal structures that shape our thoughts, attitudes, and behaviors, social categories like race, gender, and class are deeply embedded in our experiences—and often an ingrained part of our identity. But rather than reducing individuals to essentialist categories, a more appropriate approach is to consider how various identities—demographic, professional, or situational—interact, by role-playing various personas through contextualized prompting. This involves crafting character profiles that encompass a broader array of characteristics— from contextual descriptions ("I am a young tech worker living in the United States") to broader social categories (e.g., based on political leaning or personality type)—allowing for more nuanced explorations of perspectives and experiences.

For example, instead of prompting an LLM to act as a "Black woman," which may reinforce stereotypes or oversimplify identity (Sun et al., 2025), we might construct a more holistic persona by adding a specific context, such as "a young entrepreneur from Atlanta who is passionate about sustainable fashion and community development"; or by incorporating intersectional identities, such as "a young Black female tech worker navigating the challenges of a male-dominated field." These contextual descriptions incorporate identity but frame it within specific experiences, values, and contexts. Indeed, contextualized prompting has been shown to evoke distinct, diverse (Wang et al., 2025), and better-aligned responses from LLMs (Bui et al., 2025).

Identity is multifaceted and context-dependent, with varying salience for different individuals. Simulating human participants therefore risks misrepresenting the salience of various aspects of identity—reflecting the prompter's perspective or presumptions about which aspects of identity are important, rather than capturing the intersectional reality experienced by the simulated persona. It is therefore crucial, whether using contextualized prompting or not, to examine potential biases and limitations in the prompt. Sidestepping genuine engagement with marginalized communities further risks artificial inclusion (Agnew et al., 2024).

To mitigate the identity essentialization fallacy when simulating diverse perspectives, safeguards include using contextual prompting, incorporating intersectional approaches, performing diversity validation, ensuring transparency about limitations, and adopting collaborative methods. Consider several approaches:

1. **Contextual prompting**: Instead of using simple demographic labels, develop richer, context-specific prompts that incorporate multiple aspects of identity (including intersectionality), specific experiences, and environmental factors (Bui et al., 2025).
2. **Validation**: Compare LLM-generated responses across multiple prompting strategies and validate against actual human responses from the target population to identify when simulations misrepresent or stereotype particular groups (Sun et al., 2025).
3. **Transparency in limitations**: Explicitly acknowledge the limitations of identity simulation in research reports, including the risk of reinforcing stereotypes or oversimplifying complex identities. Document the specific prompting approaches used and their potential biases (Sun et al., 2025).
4. **Collaborative approach**: When studying specific cultural or identity groups, involve members of those groups in designing prompts, validating outputs, and interpreting results (Zhao et al., 2024).

**Substitution Fallacy**

LLMs can mimic certain aspects of human behavior and cognition, but purporting them as primary tools to directly reveal the human mind reflects a substitution fallacy.

A core issue arises from the temporal limitations of LLM training data. Because LLMs are trained on historical datasets with a specific cutoff date, they represent a snapshot of human knowledge, attitudes, and behaviors at that moment. Updating through retraining is infrequent and resource-intensive. This static nature restricts their capacity to capture ongoing societal changes, new social phenomena, or evolving attitudes and behaviors, such as rapidly changing views on new technologies or social movements (Zhu et al., 2025). Without real-time adaptability, previous alignments do not guarantee current applicability. Furthermore, a model's advertised knowledge cutoff often differs from its actual, or effective, knowledge cutoff. The functional knowledge of LLMs frequently corresponds to older text versions that predate the stated cutoff. This discrepancy stems from widespread temporal misalignments within large pretraining corpora—for instance, older documents lingering in recent web crawls—and from the incomplete removal of outdated or duplicated content during data processing (Cheng et al., 2024).

This fallacy persists even if we disregard challenges related to the static and historically bound nature of LLM training data—or issues of grounding, embodiment, and subjective experience. As the average human fallacy illustrates, responses from LLMs cannot be assumed *a priori* to represent average responses of the targeted human group. Even when LLMs and humans show alignment, this correlation should not be confused with equivalence in cognitive processes or mechanisms (the alignment as explanation fallacy; **Box 4**). This leads to an epistemic dilemma: Generalizing findings from LLMs to humans requires corroboration with actual human data, undermining the basic premise of the substitution proposition.

Additionally, substitution risks creating closed-loop information systems. When models trained on historical data are used as primary tools for generating new data, they perpetuate a self-referential loop that creates a distorted view of the present by amplifying the past (including its biases, errors, and oversights) rather than reflecting current human thought or behavior. This can lead to misleading statistical inferences reflecting model artifacts rather than genuine psychological phenomena and behavioral patterns.

Even with up-to-date training data, excluding human participants leaves LLMs simulating humans in ways detached from rich, evolving realities. Such detachment can entrench outdated knowledge, weaken the diversity vital for human progress, and create epistemic echo chambers. Thus, LLMs should serve as supplementary rather than primary tools for understanding the human mind.

To mitigate the substitution fallacy, safeguards include using sequential validation with human participants, benchmarking against time-sensitive data, integrating mixed methods, ensuring temporal transparency, and implementing closed-loop detection techniques. Consider several approaches:

1. **Sequential validation**: Implement a sequential research design where LLM explorations are followed by validation with human participants. Use LLMs for hypothesis generation or initial exploration, but validate key findings with relevant human data (Gui & Toubia, 2023).
2. **Benchmarking against time-sensitive data**: Regularly benchmark LLM responses against recent human data to assess temporal drift in model outputs compared to current human attitudes and behaviors. This helps establish the temporal boundaries of generalizability for LLM-based findings (Cheng et al., 2024).
3. **Temporal transparency**: Explicitly document training data cutoff dates and potential temporal limitations in research reports, particularly in rapidly evolving domains (Cheng et al., 2024).

## Concluding Remarks

Recent advances in human-level AI are renewing the classic debate on the role of computing artifacts in understanding the human mind and brain (Simon, 1983). This paper critically assesses the emerging proposition of substituting human participants with LLMs in behavioral and social sciences. By exposing six fallacies inherent in this replacement perspective, the analysis underscores that, despite their human-like language production capabilities, current LLMs do not—and as presently conceived, cannot—substitute for the nuanced intricacies of human thought. Unlike the statistical text prediction that drives current LLMs, human intelligence emerges from embodied interaction with the world—grounded in sensory experiences, enriched by multimodal integration, and shaped by subjective consciousness. The predominantly linguistic nature of LLMs further constrains their ability to capture the breadth of human experience, including nonverbal cues, implicit attitudes, and real-world behaviors.

By identifying challenges to research integrity and providing practical guidelines, the analysis supports the simulation perspective: LLMs serve as tools for simulating roles and modeling cognitive processes, complementing but not replacing humans. As outlined in **Box 4**, this perspective helps investigators distinguish between research contexts where output-level simulation suffices (pragmatic applications like rapid prototyping) and those requiring deeper mechanistic evidence (theoretical claims about cognitive processes). In practice, researchers should leverage LLMs primarily for hypothesis generation, theory development, and rapid prototyping—then validate with human participants. This sequential approach capitalizes on model strengths (comprehensive knowledge, efficient simulation) while acknowledging their limitations (lack of grounding, representational biases). Implementing the controls and considerations outlined for each fallacy can substantially improve research quality and interpretability.

As emphasized in **Box 3**, understanding model limitations requires distinguishing between technical and conceptual constraints. While technical limitations may be addressed through engineering advances, conceptual limitations represent fundamental challenges to using LLMs as psychological models. As these technologies evolve, we must continuously re-evaluate their capabilities and limitations, develop appropriate benchmarks, and establish guidelines for responsible integration. This perspective invites us to reconsider the role of AI in behavioral and cognitive science—as a mirror through which we can better understand the similarities and differences between human intelligence and machine intelligence. And the limitations of apparently human-like models in replicating human thought may bring us a deeper appreciation of the complexity and wonder of the mind.

## References

Abdurahman, S., Salkhordeh Ziabari, A., Moore, A. K., Bartels, D. M., & Dehghani, M. (2025). A primer for evaluating large language models in social-science research. *Advances in Methods and Practices in Psychological Science*, *8*(2), 25152459251325174. https://doi.org/10.1177/25152459251325174

Agnew, W., Bergman, A. S., Chien, J., Díaz, M., El-Sayed, S., Pittman, J., . . . McKee, K. R. (2024). *The illusion of artificial inclusion* Proceedings of the 2024 CHI Conference on Human Factors in Computing Systems, Honolulu, HI, USA. https://doi.org/10.1145/3613904.3642703

Aher, G. V., Arriaga, R. I., & Kalai, A. T. (2023). *Using large language models to simulate multiple humans and replicate human subject studies* Proceedings of the 40th International Conference on Machine Learning, Honolulu, Hawaii, USA. https://proceedings.mlr.press/v202/aher23a.html

Bao, G., Zhang, H., Wang, C., Yang, L., & Zhang, Y. (2024). How likely do LLMs with CoT mimic human reasoning? *arXiv:2402.16048*. https://doi.org/10.48550/arXiv.2402.16048

Bender, E. M., & Koller, A. (2020). Climbing towards NLU: On meaning, form, and understanding in the age of data. Proceedings of the 58th Annual Meeting of the Association for Computational Linguistics, Online.

Bengio, Y. (2024). *FAQ on catastrophic AI risks*. Retrieved September 8, 2024 from https://yoshuabengio.org/2023/06/24/faq-on-catastrophic-ai-risks/

Binz, M., & Schulz, E. (2023a). Using cognitive psychology to understand GPT-3. *Proceedings of the National Academy of Sciences of the United States of America*, *120*(6), e2218523120. https://doi.org/10.1073/pnas.2218523120

Binz, M., & Schulz, E. (2023b). Turning large language models into cognitive models. *arXiv:2306.03917*. https://doi.org/10.48550/arXiv.2306.03917

Block, N. (1981). Psychologism and behaviorism. *Philosophical Review*, *90*(1), 5-43. https://doi.org/10.2307/2184371

Bowers, J. S., Malhotra, G., Dujmovic, M., Llera Montero, M., Tsvetkov, C., Biscione, V., . . . Blything, R. (2023). Deep problems with neural network models of human vision. *Behavioral and Brain Sciences*, *46*, e385. https://doi.org/10.1017/S0140525X22002813

Brucks, M., & Toubia, O. (2025). Prompt architecture induces methodological artifacts in large language models. *PLOS ONE*, *20*(4), e0319159. https://doi.org/10.1371/journal.pone.0319159

Bui, N., Nguyen, H. T., Kumar, S., Theodore, J., Qiu, W., Nguyen, V. A., & Ying, R. (2025). Mixture-of-personas language models for population simulation. *arXiv:2504.05019*. https://doi.org/10.48550/arXiv.2504.05019

Butlin, P., Long, R., Elmoznino, E., Bengio, Y., Birch, J., Constant, A., . . . Ji, X. (2023). Consciousness in artificial intelligence: Insights from the science of consciousness. *arXiv:2308.08708*. https://doi.org/10.48550/arXiv.2308.08708

Cheng, J., Marone, M., Weller, O., Lawrie, D., Khashabi, D., & Van Durme, B. (2024). Dated data: Tracing knowledge cutoffs in large language models. *arXiv:2403.12958*. https://doi.org/10.48550/arXiv.2403.12958

Chuang, Y.-S., Studdiford, Z., Nirunwiroj, K., Goyal, A., Frigo, V. V., Yang, S., . . . Rogers, T. T. (2024). Beyond demographics: Aligning role-playing LLM-based agents using human belief networks. *arXiv:2406.17232*. https://doi.org/10.48550/arXiv.2406.17232

Church, A. (1936). An unsolvable problem of elementary number theory. *American Journal of Mathematics*, *58*(2), 345-363. https://doi.org/10.2307/2371045

Colombatto, C., & Fleming, S. M. (2024). Folk psychological attributions of consciousness to large language models. *Neuroscience of Consciousness*, *2024*(1), niae013. https://doi.org/10.1093/nc/niae013

Cook, T. D., & Campbell, D. T. (1979). *Quasi-experimentation: Design & analysis issues for field settings*. Houghton Mifflin.

Crockett, M., & Messeri, L. (2023). Should large language models replace human participants? *PsyArXiv*. https://doi.org/10.31234/osf.io/4zdx9

Cronbach, L. J., & Meehl, P. E. (1955). Construct validity in psychological tests. *Psychological Bulletin*, *52*(4), 281-302. https://doi.org/10.1037/h0040957

Dasgupta, I., Lampinen, A. K., Chan, S. C., Sheahan, H. R., Creswell, A., Kumaran, D., . . . Hill, F. (2022). Language models show human-like content effects on reasoning tasks. *arXiv:2207.07051*. https://doi.org/10.48550/arXiv.2207.07051

Demszky, D., Yang, D., Yeager, D. S., Bryan, C. J., Clapper, M., Chandhok, S., . . . Pennebaker, J. W. (2023). Using large language models in psychology. *Nature Reviews Psychology*, *2*(11), 688-701. https://doi.org/10.1038/s44159-023-00241-5

Dettki, H. M., Lake, B. M., Wu, C. M., & Rehder, B. (2025). Do large language models reason causally like us? Even better? *arXiv:2502.10215*. https://doi.org/10.48550/arXiv.2502.10215

Dillion, D., Tandon, N., Gu, Y., & Gray, K. (2023). Can AI language models replace human participants? *Trends in Cognitive Sciences*, *27*(7), 597-600. https://doi.org/10.1016/j.tics.2023.04.008

Doerig, A., Sommers, R. P., Seeliger, K., Richards, B., Ismael, J., Lindsay, G. W., . . . Kietzmann, T. C. (2023). The neuroconnectionist research programme. *Nature Reviews Neuroscience*, *24*(7), 431-450. https://doi.org/10.1038/s41583-023-00705-w

Ferreira, F., Bailey, K. G. D., & Ferraro, V. (2002). Good-enough representations in language comprehension. *Current Directions in Psychological Science*, *11*(1), 11-15. https://doi.org/10.1111/1467-8721.00158

Firestone, C. (2020). Performance vs. competence in human–machine comparisons. *Proceedings of the National Academy of Sciences of the United States of America*, *117*(43), 26562-26571. https://doi.org/10.1073/pnas.1905334117

Fokas, A. S. (2023). Can artificial intelligence reach human thought? *PNAS Nexus*, *2*(12), pgad409. https://doi.org/10.1093/pnasnexus/pgad409

Gao, Y., Lee, D., Burtch, G., & Fazelpour, S. (2024). Take caution in using LLMs as human surrogates: Scylla ex machina. *arXiv:2410.19599*. https://doi.org/10.48550/arXiv.2410.19599

Grossmann, I., Feinberg, M., Parker, D. C., Christakis, N. A., Tetlock, P. E., & Cunningham, W. A. (2023). AI and the transformation of social science research. *Science*, *380*(6650), 1108-1109. https://doi.org/10.1126/science.adi1778

Guest, O., & Martin, A. E. (2023). On logical inference over brains, behaviour, and artificial neural networks. *Computational Brain & Behavior*, *6*(2), 213-227. https://doi.org/10.1007/s42113-022-00166-x

Gui, G., & Toubia, O. (2023). The challenge of using LLMs to simulate human behavior: A causal inference perspective. *arXiv:2312.15524*. https://doi.org/10.48550/arXiv.2312.15524

Gurnee, W., & Tegmark, M. (2023). Language models represent space and time. *arXiv:2310.02207*. https://doi.org/10.48550/arXiv.2310.02207

GX-Chen, A., Lin, D., Samiei, M., Precup, D., Richards, B. A., Fergus, R., & Marino, K. (2025). Language agents mirror human causal reasoning biases: How can we help them think like scientists? *arXiv:2505.09614*. https://doi.org/10.48550/arXiv.2505.09614

Hagendorff, T., Fabi, S., & Kosinski, M. (2023). Human-like intuitive behavior and reasoning biases emerged in large language models but disappeared in ChatGPT. *Nature Computational Science*, *3*(10), 833-838. https://doi.org/10.1038/s43588-023-00527-x

Harding, J., D'Alessandro, W., Laskowski, N. G., & Long, R. (2024). AI language models cannot replace human research participants. *AI & Society*, *39*, 2603–2605. https://doi.org/10.1007/s00146-023-01725-x

Harding, J., & Sharadin, N. (in press). What is it for a machine learning model to have a capability? *The British Journal for the Philosophy of Science*. https://doi.org/10.1086/732153

Hu, J., Mahowald, K., Lupyan, G., Ivanova, A., & Levy, R. (2024). Language models align with human judgments on key grammatical constructions. *Proceedings of the National Academy of Sciences of the United States of America*, *121*(36), e2400917121. https://doi.org/10.1073/pnas.2400917121

Hu, T., Kyrychenko, Y., Rathje, S., Collier, N., van der Linden, S., & Roozenbeek, J. (2025). Generative language models exhibit social identity biases. *Nature Computational Science*, *5*(1), 65-75. https://doi.org/10.1038/s43588-024-00741-1

Huang, P.-H., Lin, Z., Imbot, S., Fu, W., & Tu, E. (2025). Analysis of LLM bias (Chinese propaganda and anti-US sentiment) in DeepSeek-R1 vs. ChatGPT o3-mini-high. *arXiv:2506.01814*. https://doi.org/10.48550/arXiv.2506.01814

Ibrahim, L., & Cheng, M. (2025). Thinking beyond the anthropomorphic paradigm benefits LLM research. *arXiv:2502.09192*. https://doi.org/10.48550/arXiv.2502.09192

Ivanova, A. A. (2025). How to evaluate the cognitive abilities of LLMs. *Nature Human Behaviour*, *9*(2), 230-233. https://doi.org/10.1038/s41562-024-02096-z

Jackson, F. (1982). Epiphenomenal qualia. *The Philosophical Quarterly*, *32*(127), 127-136. https://doi.org/10.2307/2960077

Jackson, F. (1986). What Mary didn't know. *The Journal of Philosophy*, *83*(5), 291-295. https://doi.org/10.2307/2026143

Jin, Z., Kleiman-Weiner, M., Piatti, G., Levine, S., Liu, J., Gonzalez, F., . . . Mihalcea, R. (2024). Language model alignment in multilingual trolley problems. *arXiv:2407.02273*. https://doi.org/10.48550/arXiv.2407.02273

Jones, C. R., & Bergen, B. (2024). Does word knowledge account for the effect of world knowledge on pronoun interpretation? *Language and Cognition*, 1-32. https://doi.org/10.1017/langcog.2024.2

Jones, C. R., Chang, T. A., Coulson, S., Michaelov, J. A., Trott, S., & Bergen, B. (2022). Distributional semantics still can't account for affordances. Proceedings of the Annual Meeting of the Cognitive Science Society, Retrieved from https://escholarship.org/uc/item/44z7r3j3.

Kamoi, R., Das, S. S. S., Lou, R., Ahn, J. J., Zhao, Y., Lu, X., . . . Vummanthala, S. R. (2024). Evaluating LLMs at detecting errors in LLM responses. *arXiv:2404.03602*. https://doi.org/10.48550/arXiv.2404.03602

Kerschbaumer, S., Voracek, M., Aczél, B., Anderson, S. F., Booth, B. M., Buchanan, E. M., . . . Tran, U. S. (2025). VALID: A checklist-based approach for improving validity in psychological research. *Advances in Methods and Practices in Psychological Science*, *8*(1), 25152459241306432. https://doi.org/10.1177/25152459241306432

Kim, J. S., Elli, G. V., & Bedny, M. (2019). Knowledge of animal appearance among sighted and blind adults. *Proceedings of the National Academy of Sciences of the United States of America*, *116*(23), 11213-11222. https://doi.org/10.1073/pnas.1900952116

Kozlowski, A. C., & Evans, J. A. (2024). Simulating subjects: The promise and peril of AI stand-ins for social agents and interactions. *SocArXiv*. https://doi.org/10.31235/osf.io/vp3j2

Lahoti, P., Blumm, N., Ma, X., Kotikalapudi, R., Potluri, S., Tan, Q., . . . Beutel, A. (2023). Improving diversity of demographic representation in large language models via collective-critiques and self-voting. *arXiv:2310.16523*. https://doi.org/10.48550/arXiv.2310.16523

Lake, B. M., & Murphy, G. L. (2023). Word meaning in minds and machines. *Psychological Review*, *130*(2), 401-431. https://doi.org/10.1037/rev0000297

Lee, M. H., Montgomery, J. M., & Lai, C. K. (2024). Large language models portray socially subordinate groups as more homogeneous, consistent with a bias observed in humans. Proceedings of the 2024 ACM Conference on Fairness, Accountability, and Transparency, Rio de Janeiro, Brazil.

Lee, S., Li, M., Lai, B., Jia, W., Ryan, F., Cao, X., . . . Yang, D. (2024). Towards social AI: A survey on understanding social interactions. *arXiv:2409.15316*. https://doi.org/10.48550/arXiv.2409.15316

Leivada, E., Marcus, G., Günther, F., & Murphy, E. (2023). A sentence is worth a thousand pictures: Can large language models understand hum4n l4ngu4ge and the w0rld behind w0rds? *arXiv:2308.00109*. https://doi.org/10.48550/arXiv.2308.00109

Lewis, M., Zettersten, M., & Lupyan, G. (2019). Distributional semantics as a source of visual knowledge. *Proceedings of the National Academy of Sciences of the United States of America*, *116*(39), 19237-19238. https://doi.org/10.1073/pnas.1910148116

Li, C., & Qi, Y. (2025). Toward accurate psychological simulations: Investigating LLMs' responses to personality and cultural variables. *Computers in Human Behavior*, *170*, 108687. https://doi.org/10.1016/j.chb.2025.108687

Li, Y., Huang, Y., Wang, H., Zhang, X., Zou, J., & Sun, L. (2024). Quantifying AI psychology: A psychometrics benchmark for large language models. *arXiv:2406.17675*. https://doi.org/10.48550/arXiv.2406.17675

Lin, Z. (2023). Why and how to embrace AI such as ChatGPT in your academic life. *Royal Society Open Science*, *10*, 230658. https://doi.org/10.1098/rsos.230658

Lin, Z. (2024). How to write effective prompts for large language models. *Nature Human Behaviour*, *8*(4), 611-615. https://doi.org/10.1038/s41562-024-01847-2

Lin, Z. (2025). Beyond principlism: Practical strategies for ethical AI use in research practices. *AI and Ethics*, *5*, 2719–2731. https://doi.org/10.1007/s43681-024-00585-5

Lin, Z., & Li, N. (2023). Global diversity of authors, editors, and journal ownership across subdisciplines of psychology: Current state and policy implications. *Perspectives on Psychological Science*, *18*(2), 358-377. https://doi.org/10.1177/17456916221091831

Mahowald, K., Ivanova, A. A., Blank, I. A., Kanwisher, N., Tenenbaum, J. B., & Fedorenko, E. (2024). Dissociating language and thought in large language models. *Trends in Cognitive Sciences*, *28*(6), 517-540. https://doi.org/10.1016/j.tics.2024.01.011

Manning, C. D., Clark, K., Hewitt, J., Khandelwal, U., & Levy, O. (2020). Emergent linguistic structure in artificial neural networks trained by self-supervision. *Proceedings of the National Academy of Sciences of the United States of America*, *117*(48), 30046-30054. https://doi.org/10.1073/pnas.1907367117

Marjieh, R., Sucholutsky, I., van Rijn, P., Jacoby, N., & Griffiths, T. L. (2024). Large language models predict human sensory judgments across six modalities. *Scientific Reports*, *14*(1), 21445. https://doi.org/10.1038/s41598-024-72071-1

McCoy, R. T., Yao, S., Friedman, D., Hardy, M. D., & Griffiths, T. L. (2024). Embers of autoregression show how large language models are shaped by the problem they are trained to solve. *Proceedings of the National Academy of Sciences of the United States of America*, *121*(41), e2322420121. https://doi.org/10.1073/pnas.2322420121

Millière, R. (2024). Language models as models of language. *arXiv:2408.07144*. https://doi.org/10.48550/arXiv.2408.07144

Millière, R., & Buckner, C. (2024a). A philosophical introduction to language models – Part I: Continuity with classic debates. *arXiv:2401.03910*. https://doi.org/10.48550/arXiv.2401.03910

Millière, R., & Buckner, C. (2024b). A philosophical introduction to language models – Part II: The way forward. *arXiv:2405.03207*. https://doi.org/10.48550/arXiv.2405.03207

Mirzadeh, I., Alizadeh, K., Shahrokhi, H., Tuzel, O., Bengio, S., & Farajtabar, M. (2024). GSM-Symbolic: Understanding the limitations of mathematical reasoning in large language models. *arXiv:2410.05229*. https://doi.org/10.48550/arXiv.2410.05229

Mitchell, M. (2024). Debates on the nature of artificial general intelligence. *Science*, *383*(6689), eado7069. https://doi.org/10.1126/science.ado7069

Mitchell, M., & Krakauer, D. C. (2023). The debate over understanding in AI's large language models. *Proceedings of the National Academy of Sciences of the United States of America*, *120*(13), e2215907120. https://doi.org/10.1073/pnas.2215907120

Momennejad, I., Hasanbeig, H., Vieira Frujeri, F., Sharma, H., Jojic, N., Palangi, H., . . . Larson, J. (2023). Evaluating cognitive maps and planning in large language models with cogeval. *Advances in Neural Information Processing Systems*, *36*, 69736-69751.

Mozikov, M., Severin, N., Bodishtianu, V., Glushanina, M., Baklashkin, M., Savchenko, A. V., & Makarov, I. (2024). The good, the bad, and the Hulk-like GPT: Analyzing emotional decisions of large language models in cooperation and bargaining games. *arXiv:2406.03299*. https://doi.org/10.48550/arXiv.2406.03299

Murthy, S. K., Ullman, T., & Hu, J. (2024). One fish, two fish, but not the whole sea: Alignment reduces language models' conceptual diversity. *arXiv:2411.04427*. https://doi.org/10.48550/arXiv.2411.04427

Namboodiripad, S., Kutlu, E., Babel, A., Baese-Berk, M., Block, A., Carlson, M. T., . . . Wright, K. E. (2023). Essentialist characterizations of language are an obstacle to accuracy, progress, and justice in science. *PsyArXiv*. https://doi.org/10.31234/osf.io/jn3ct

Niszczota, P., Janczak, M., & Misiak, M. (2025). Large language models can replicate cross-cultural differences in personality. *Journal of Research in Personality*, *115*, 104584. https://doi.org/10.1016/j.jrp.2025.104584

Oh, S., & Demberg, V. (2025). Robustness of large language models in moral judgements. *Royal Society Open Science*, *12*(4), 241229. https://doi.org/10.1098/rsos.241229

OpenAI. (2024). *OpenAI charter*. Retrieved September 8, 2024 from https://openai.com/charter/

Park, P. S., Schoenegger, P., & Zhu, C. (2024). Diminished diversity-of-thought in a standard large language model. *Behavior Research Methods*. https://doi.org/10.3758/s13428-023-02307-x

Pellert, M., Lechner, C. M., Wagner, C., Rammstedt, B., & Strohmaier, M. (2024). AI psychometrics: Assessing the psychological profiles of large language models through psychometric inventories. *Perspectives on Psychological Science*, *19*(5), 808-826. https://doi.org/10.1177/17456916231214460

Prentice, D. A., & Miller, D. T. (2006). Essentializing differences between women and men. *Psychological Science*, *17*(2), 129-135. https://doi.org/10.1111/j.1467-9280.2006.01675.x

Qu, Y., Du, P., Che, W., Wei, C., Zhang, C., Ouyang, W., . . . Liu, Q. (2024). Promoting interactions between cognitive science and large language models. *The Innovation*, *5*(2), 100579. https://doi.org/10.1016/j.xinn.2024.100579

Richards, B. A., Lillicrap, T. P., Beaudoin, P., Bengio, Y., Bogacz, R., Christensen, A., . . . Kording, K. P. (2019). A deep learning framework for neuroscience. *Nature Neuroscience*, *22*(11), 1761-1770. https://doi.org/10.1038/s41593-019-0520-2

Santurkar, S., Durmus, E., Ladhak, F., Lee, C., Liang, P., & Hashimoto, T. (2023). *Whose opinions do language models reflect?* Proceedings of the 40th International Conference on Machine Learning, Proceedings of Machine Learning Research. https://proceedings.mlr.press/v202/santurkar23a.html

Sarstedt, M., Adler, S. J., Rau, L., & Schmitt, B. (2024). Using large language models to generate silicon samples in consumer and marketing research: Challenges, opportunities, and guidelines. *Psychology & Marketing*, *41*, 1254–1270. https://doi.org/10.1002/mar.21982

Schmidt, A., Elagroudy, P., Draxler, F., Kreuter, F., & Welsch, R. (2024). Simulating the human in HCD with ChatGPT: Redesigning interaction design with AI. *Interactions*, *31*(1), 24–31. https://doi.org/10.1145/3637436

Searle, J. R. (1980). Minds, brains, and programs. *Behavioral and Brain Sciences*, *3*(3), 417-424. https://doi.org/10.1017/S0140525X00005756

Shanahan, M. (2024). Talking about large language models. *Communications of the ACM*, *67*(2), 68–79. https://doi.org/10.1145/3624724

Shanahan, M., McDonell, K., & Reynolds, L. (2023). Role play with large language models. *Nature*, *623*(7987), 493-498. https://doi.org/10.1038/s41586-023-06647-8

Shardlow, M., & Przybyla, P. (2024). Deanthropomorphising NLP: Can a language model be conscious? *PLOS ONE*, *19*(12), e0307521. https://doi.org/10.1371/journal.pone.0307521

Shiffrin, R., & Mitchell, M. (2023). Probing the psychology of AI models. *Proceedings of the National Academy of Sciences of the United States of America*, *120*(10), e2300963120. https://doi.org/10.1073/pnas.2300963120

Simon, H. A. (1983). Why should machines learn? In R. S. Michalski, J. G. Carbonell, & T. M. Mitchell (Eds.), *Machine Learning* (pp. 25-37). Morgan Kaufmann. https://doi.org/10.1016/B978-0-08-051054-5.50006-6

Stanley, J. C., & Campbell, D. T. (1963). *Experimental and quasi-experimental designs for research*. Rand McNally.

Strachan, J. W. A., Albergo, D., Borghini, G., Pansardi, O., Scaliti, E., Gupta, S., . . . Becchio, C. (2024). Testing theory of mind in large language models and humans. *Nature Human Behaviour*, *8*(7), 1285-1295. https://doi.org/10.1038/s41562-024-01882-z

Suh, J., Jahanparast, E., Moon, S., Kang, M., & Chang, S. (2025). Language model fine-tuning on scaled survey data for predicting distributions of public opinions. *arXiv:2502.16761*. https://doi.org/10.48550/arXiv.2502.16761

Sun, H., Pei, J., Choi, M., & Jurgens, D. (2025). Sociodemographic prompting is not yet an effective approach for simulating subjective judgments with LLMs. Proceedings of the 2025 Conference of the Nations of the Americas Chapter of the Association for Computational Linguistics: Human Language Technologies (Volume 2: Short Papers), Albuquerque, New Mexico.

Suri, G., Slater, L. R., Ziaee, A., & Nguyen, M. (2024). Do large language models show decision heuristics similar to humans? A case study using GPT-3.5. *Journal of Experimental Psychology: General*, *153*(4), 1066-1075. https://doi.org/10.1037/xge0001547

Tao, Y., Viberg, O., Baker, R. S., & Kizilcec, R. F. (2024). Cultural bias and cultural alignment of large language models. *PNAS Nexus*, *3*(9). https://doi.org/10.1093/pnasnexus/pgae346

Tjuatja, L., Chen, V., Wu, T., Talwalkwar, A., & Neubig, G. (2024). Do LLMs exhibit human-like response biases? A case study in survey design. *Transactions of the Association for Computational Linguistics*, *12*, 1011-1026. https://doi.org/10.1162/tacl_a_00685

Vallor, S. (2024). *The AI mirror: How to reclaim our humanity in an age of machine thinking*. Oxford University Press.

Van Rooij, I., Guest, O., Adolfi, F. G., de Haan, R., Kolokolova, A., & Rich, P. (2024). Reclaiming AI as a theoretical tool for cognitive science. *Computational Brain & Behavior*, *7*, 616-636. https://doi.org/10.1007/542113-024-00217-5

Vaswani, A., Shazeer, N., Parmar, N., Uszkoreit, J., Jones, L., Gomez, A. N., . . . Polosukhin, I. (2017). *Attention is all you need* Proceedings of the 31st International Conference on Neural Information Processing Systems, Long Beach, California, USA.

Wang, A., Morgenstern, J., & Dickerson, J. P. (2025). Large language models that replace human participants can harmfully misportray and flatten identity groups. *Nature Machine Intelligence*, *7*(3), 400-411. https://doi.org/10.1038/s42256-025-00986-z

Wang, X., Li, X., Yin, Z., Wu, Y., & Liu, J. (2023). Emotional intelligence of large language models. *Journal of Pacific Rim Psychology*, *17*, 18344909231213958. https://doi.org/10.1177/18344909231213958

Webb, T., Holyoak, K. J., & Lu, H. (2023). Emergent analogical reasoning in large language models. *Nature Human Behaviour*, *7*(9), 1526-1541. https://doi.org/10.1038/s41562-023-01659-w

Wicke, P., & Wachowiak, L. (2024, August). Exploring spatial schema intuitions in large language and vision models. In L.-W. Ku, A. Martins, & V. Srikumar, *Findings of the Association for Computational Linguistics: ACL 2024* Bangkok, Thailand.

Xu, Q., Peng, Y., Nastase, S. A., Chodorow, M., Wu, M., & Li, P. (2025). Large language models without grounding recover non-sensorimotor but not sensorimotor features of human concepts. *Nature Human Behaviour*. https://doi.org/10.1038/s41562-025-02203-8

Yax, N., Anllo, H., & Palminteri, S. (2024). Studying and improving reasoning in humans and machines. *Communications Psychology*, *2*(1), 51. https://doi.org/10.1038/s44271-024-00091-8

Ye, A., Moore, J., Novick, R., & Zhang, A. X. (2024). Language models as critical thinking tools: A case study of philosophers. *arXiv:2404.04516*. https://doi.org/10.48550/arXiv.2404.04516

Yildirim, I., & Paul, L. A. (2024). From task structures to world models: What do LLMs know? *Trends in Cognitive Sciences*, *28*(5), 404-415. https://doi.org/10.1016/j.tics.2024.02.008

Zan, M., Zhang, Y., Zhang, B., Liu, F., & Cheng, D. (2025). Can the capability of large language models be described by human ability? A meta study. *arXiv:2504.12332*. https://doi.org/10.48550/arXiv.2504.12332

Zhao, X., Sriwarnasinghe, S. M., Tang, J., Wang, S., Wang, H., & Morikawa, S. (2024). Collaborative participatory research with LLM agents in South Asia: An empirically-grounded methodological initiative and agenda from field evidence in Sri Lanka. *arXiv:2411.08294*. https://doi.org/10.48550/arXiv.2411.08294

Zhou, L., Schellaert, W., Martinez-Plumed, F., Moros-Daval, Y., Ferri, C., & Hernandez-Orallo, J. (2024). Larger and more instructable language models become less reliable. *Nature*. https://doi.org/10.1038/s41586-024-07930-y

Zhu, C., Chen, N., Gao, Y., Zhang, Y., Tiwari, P., & Wang, B. (2025). Is your LLM outdated? A deep look at temporal generalization. Proceedings of the 2025 Conference of the Nations of the Americas Chapter of the Association for Computational Linguistics: Human Language Technologies (Volume 1: Long Papers), Albuquerque, New Mexico.

Ziems, C., Held, W., Shaikh, O., Chen, J., Zhang, Z., & Yang, D. (2024). Can large language models transform computational social science? *Computational Linguistics*, *50*(1), 237-291. https://doi.org/10.1162/coli_a_00502